%% file: main.tex
\documentclass[runningheads,bookmarksdepth=2]{llncs}

\usepackage{iflang}

\usepackage[T1]{fontenc}

\usepackage{mdframed}

\usepackage{graphicx}

\usepackage[svgnames,dvipsnames]{xcolor}

\usepackage{enumitem}

\usepackage{xspace}

\usepackage{tikz}
\usetikzlibrary{arrows.meta}

\usepackage{hyperref}
\hypersetup{
  bookmarksopen=true,    % Show bookmarks bar?
  pdftitle={Exploring Micro Frontends:\break A Case Study Application in E-Commerce},
  pdfauthor={R. Kojo el al.},
  pdfcreator={pdflatex}, % Creator of the document
  pdfnewwindow=true,     % Links in new window
  colorlinks=true,       % false: boxed links; true: colored links
  linkcolor=blue,        % Color of internal links
  citecolor=DarkGreen,   % Color of links to bibliography
  filecolor=green,       % Color of file links
  urlcolor=DarkRed       % Color of external links
}

\usepackage[nameinlink,noabbrev]{cleveref}

\newcommand{\etal}{\textit{et al.}\xspace}

\newcommand{\Quote}[1]{``\textit{#1}''\xspace}

\newcommand{\Pattern}[1]{\textsc{#1}\xspace}

\newcommand{\Team}[1]{\textbf{#1}\xspace}
\newcommand{\Company}{\Team{Company~A}}

\newcommand{\Component}[1]{\texttt{#1}\xspace}
\newcommand{\Marketplace}{\Component{Marketplace}}
\newcommand{\Aquarelle}{\Component{Aquarelle}}
\newcommand{\Brindle}{\Component{Brindle}}
\newcommand{\BrindleCore}{\Component{Brindle Core}}
\newcommand{\BrindleDiscovery}{\Component{Brindle Discovery}}
\newcommand{\Fortelle}{\Component{Fortelle}}
\newcommand{\Nuvelle}{\Component{Nuvelle}}
\newcommand{\Veloura}{\Component{Veloura}}

\begin{document}

%%%%%%%%%%%%%%%%%%%%%%%%%%%%%%%%%%%%%%%%%%%%%%%%%%%%%%%%%%%%%%%%%%%%%%%%%%%%%%%%
%                                    TITLE                                     %
%%%%%%%%%%%%%%%%%%%%%%%%%%%%%%%%%%%%%%%%%%%%%%%%%%%%%%%%%%%%%%%%%%%%%%%%%%%%%%%%

% Title in the first page (breaking line after :)
\title{Exploring Micro Frontends:\break A Case Study Application in E-Commerce}
% Title in the header (no line breaks)
\titlerunning{Exploring Micro Frontends: A Case Study Application in E-Commerce}

%%%%%%%%%%%%%%%%%%%%%%%%%%%%%%%%%%%%%%%%%%%%%%%%%%%%%%%%%%%%%%%%%%%%%%%%%%%%%%%%
%                                   AUTHORS                                    %
%%%%%%%%%%%%%%%%%%%%%%%%%%%%%%%%%%%%%%%%%%%%%%%%%%%%%%%%%%%%%%%%%%%%%%%%%%%%%%%%

\author{
Ricardo Hideki Hangai Kojo\inst{1}\orcidID{0009-0007-5614-9671} \and \\
Luiz Fernando Corte Real\inst{1}\orcidID{0009-0003-1086-0602} \and \\
Renato Cordeiro Ferreira\inst{1,2}\orcidID{0000-0001-7296-7091} \and \\
Thatiane de Oliveira Rosa\inst{1,3}\orcidID{0000-0002-3980-0051} \and \\
Alfredo Goldman\inst{1}\orcidID{0000-0001-5746-4154}
}

\authorrunning{R. Kojo \etal}

%%%%%%%%%%%%%%%%%%%%%%%%%%%%%%%%%%%%%%%%%%%%%%%%%%%%%%%%%%%%%%%%%%%%%%%%%%%%%%%%
%                                 AFFILIATIONS                                 %
%%%%%%%%%%%%%%%%%%%%%%%%%%%%%%%%%%%%%%%%%%%%%%%%%%%%%%%%%%%%%%%%%%%%%%%%%%%%%%%%

\institute{%
Mathematics and Statistics Institute (IME), University of São Paulo (USP),
Brazil \and
Jheronimus Academy of Data Science (JADS), Tilburg University~(TiU) and
Technical University of Eindhoven~(TUe), The Netherlands \and
Federal Institute of Tocantins (IFTO), Brazil \and
\email{\{renatocf,gold\}@ime.usp.br} \\
\email{thatiane@ifto.edu.br}
}

%%%%%%%%%%%%%%%%%%%%%%%%%%%%%%%%%%%%%%%%%%%%%%%%%%%%%%%%%%%%%%%%%%%%%%%%%%%%%%%%
%                                    TITLE                                     %
%%%%%%%%%%%%%%%%%%%%%%%%%%%%%%%%%%%%%%%%%%%%%%%%%%%%%%%%%%%%%%%%%%%%%%%%%%%%%%%%

% typeset the header of the contribution
\maketitle 

%%%%%%%%%%%%%%%%%%%%%%%%%%%%%%%%%%%%%%%%%%%%%%%%%%%%%%%%%%%%%%%%%%%%%%%%%%%%%%%%
%                                  ABSTRACT                                    %
%%%%%%%%%%%%%%%%%%%%%%%%%%%%%%%%%%%%%%%%%%%%%%%%%%%%%%%%%%%%%%%%%%%%%%%%%%%%%%%%

\begin{abstract}
%% Motivation
In the micro frontends architectural style, the frontend is divided into smaller
components, which can range from a simple button to an entire page. The goal is
to improve scalability, resilience, and team independence, albeit at the cost of
increased complexity and infrastructure demands. 
%% Objective
This paper seeks to provide insights into when the adoption of micro frontends 
may be worthwhile, particularly in an industry context, considering that research 
in this area is still evolving.
%% Methodology
To achieve this, we conducted an investigation into the state of the art of
micro frontends, based on both academic and gray literature.
We then implemented this architectural style in a marketplace for handcrafted
products, which already used microservices.
Finally, we evaluated the implementation through a semi-open questionnaire
with the developers.
%% Justification
At the studied marketplace company, the need for architectural change arose
due to the tight coupling between their main system (a Java monolith) and a
dedicated frontend system. Additionally, there were deprecated technologies
and poor developer experience.
%% Execution
To address these issues, the micro frontends architecture was adopted,
along with the \Pattern{API Gateway} and \Pattern{Backend for Frontend}
patterns, and technologies such as Svelte and Fastify.
%% Analysis
Although the adoption of Micro Frontends was successful, it was not strictly
necessary to meet the company's needs. According to the analysis of the mixed
questionnaire responses, other alternatives, such as a monolithic frontend,
could have achieved comparable results.
What made adopting micro frontends the most convenient choice in the company's
context was the monolith strangulation and microservices adoption, which
facilitated implementation through infrastructure reuse and knowledge sharing
between teams.


\end{abstract}

%%%%%%%%%%%%%%%%%%%%%%%%%%%%%%%%%%%%%%%%%%%%%%%%%%%%%%%%%%%%%%%%%%%%%%%%%%%%%%%%
%                                   CHAPTERS                                   %
%%%%%%%%%%%%%%%%%%%%%%%%%%%%%%%%%%%%%%%%%%%%%%%%%%%%%%%%%%%%%%%%%%%%%%%%%%%%%%%%

\input{sections/01-introduction}
\input{sections/02-background}
\input{sections/03-architecture_overview}
\input{sections/04-implementing_micro_frontends}
\input{sections/05-current_architecture}
\input{sections/06-survey_with_devs}
\input{sections/07-related_work}
\input{sections/08-conclusion}

%%%%%%%%%%%%%%%%%%%%%%%%%%%%%%%%%%%%%%%%%%%%%%%%%%%%%%%%%%%%%%%%%%%%%%%%%%%%%%%%
%                                 BIBLIOGRAPHY                                 %
%%%%%%%%%%%%%%%%%%%%%%%%%%%%%%%%%%%%%%%%%%%%%%%%%%%%%%%%%%%%%%%%%%%%%%%%%%%%%%%%

\bibliographystyle{splncs04}
\bibliography{bibliography}

\end{document}

%% file: sections/01-introduction.tex
\section{Introduction}
\label{sec:introduction}
%%%%%%%%%%%%%%%%%%%%%%%%%%%%%%%%%%%%%%%%%%%%%%%%%%%%%%%%%%%%%%%%%%%%%%%%%%%%%%%%

The Microservices architecture, which emerged around 2012~\cite{lewisfowler14:ms},
has grown in popularity in both academia and industry, being adopted by
companies such as Amazon, Netflix, and Uber~\cite{richardson:msp}. This
architecture consists of breaking down an application into small, independent
services that work together. The loose coupling between these services brings
benefits such as maintainability, scalability, and resilience, key
characteristics of large systems that need to be reliable, always available, and
minimally prone to failure. However, this approach focuses solely on the backend
of the application, leaving the frontend aside~\cite{peltonen21:mvlr}.

In this context, the micro frontend architecture has emerged, aiming to
apply microservices concepts to the presentation layer of applications%
~\cite{peltonen21:mvlr}.
In this approach, the frontend system is divided into small parts, which may
range from a simple button to a complete page.
These parts are composed either on
  the client side (e.g., in the user's browser),
  the edge side (e.g., a CDN or Content Delivery Network), or
  the server side (e.g., a cloud server),
resulting in a unified and cohesive output~\cite{mezzalira:bmf}.

There is significant overlap between the benefits and challenges of
microservices and micro frontends. Therefore, it is reasonable to consider
adopting both architectures, leveraging existing knowledge and infrastructure.
However, the concept of micro frontends is relatively recent (introduced in
2016~\cite{geers20:mfia}), and as an emerging architectural style, the number
of academic studies on the topic is still limited, indicating that research
in this area is ongoing and evolving. Unlike microservices, which benefit 
from well-established technologies and design patterns, micro frontends still
have limited tool support. Consequently, many opportunities remain to be explored.

Thus, this study provides an empirical account of the application of the micro frontend
architectural style in an industry context, with the aim of helping understand under
which circumstances its adoption may be worthwhile.

From this, the following research questions were defined:
\begin{itemize}
    \item \textit{\textbf{RQ1:} What are the motivations and challenges involved in adopting
    a micro frontend architecture in the studied company, which already uses microservices?}
    \item \textit{\textbf{RQ2:} What are the perceived benefits and drawbacks reported by 
    developers involved in the migration from a monolithic architecture to micro frontends?}
\end{itemize}

To support the proposed objective and answer the research questions, an initial exploratory
investigation was conducted to identify relevant publications and practical insights on the
architecture. This involved consulting both academic sources (such as articles and books) 
and gray literature, including blog posts and recorded talks. The goal was not to perform
a systematic review, but to gather a representative sample of materials that could inform
the implementation and guide the study. This analysis helped identify recurring concepts,
key authors, and gaps in the existing knowledge. In addition to the literature review, 
the micro frontends architecture was implemented at Company A (a fictitious name), a 
marketplace platform that already used microservices. This objective also included 
documenting and evaluating the implementation: understanding the reasons behind the 
architectural change, outlining the development history, including the new systems, 
and explaining the rationale for the decisions made. Finally, the impacts of adopting 
micro frontends were analyzed through a questionnaire applied to the involved team members.

%% file: sections/02-background.tex
\section{Background}
\label{sec:background}
%%%%%%%%%%%%%%%%%%%%%%%%%%%%%%%%%%%%%%%%%%%%%%%%%%%%%%%%%%%%%%%%%%%%%%%%%%%%%%%%

This section aims to present the essential concepts necessary for a
comprehensive understanding of this study, assuming a basic knowledge of
software architecture and architectural styles, especially micro frontends.

\subsection{Software Architecture and Architectural Styles}
\label{subsec:software_architecture}
%%%%%%%%%%%%%%%%%%%%%%%%%%%%%%%%%%%%%%%%%%%%%%%%%%%%%%%%%%%%%%%%%%%%%%%%%%%%%%%%

Software architecture can be defined as \Quote{the set of structures needed
to reason about the system, which comprise software elements, relations among
them, and properties of both}~\cite{bassclementskazman03:sap}. It encompasses
the foundations, properties, rules, and constraints that guide system design.

The purpose of defining a software architecture is to facilitate aspects such as
development, deployment, operation, and maintenance~\cite{martin:ca}, in
addition to supporting the analysis of qualitative attributes such as
scalability, resilience, security, and maintainability. Architectural decisions
are therefore essential to the system's evolution. To identify the most suitable
architecture for addressing specific quality attributes, various architectural
styles and patterns are adopted~\cite{richards:sap}.

An architectural style prescribes a constrained set of elements
and relationships that can be used to define a system's structure%
~\cite{richards:sap}. A system that conforms to such constraints can be
considered an instance of that architectural style. It is important to note that
a single application may incorporate multiple styles.

When choosing a specific architectural style, it is crucial to consider
the motivations behind the decision, its expected benefits, and potential
challenges. Architectural decisions tend to have significant implications,
and as Brooks famously stated~\cite{brooks95:mmm}, \Quote{there are no silver
bullets in software development}; no single architectural style offers universal
solutions or only advantages.

This study explores the monolithic, microservices-based, and micro frontend
architectural styles, which are particularly relevant for understanding the
proposed research.

\vspace{-0.5em}
\subsection{Monolithic vs. Microservices-based Architectural Styles}
\label{subsec:monolith_vs_microservices}
%%%%%%%%%%%%%%%%%%%%%%%%%%%%%%%%%%%%%%%%%%%%%%%%%%%%%%%%%%%%%%%%%%%%%%%%%%%%%%%%

In a monolithic architecture, the entire application is structured in a single
layer containing all necessary logic for full system functionality, typically
resulting in a single executable artifact. This architecture is simple, making
it a common choice for new projects. However, as the team and codebase grow,
technical and organizational issues may emerge, such as maintenance
difficulties, increased bug count, slow feature development, and scalability
limitations. In such contexts, a microservices architecture may serve as a
viable alternative~\cite{richardson19:monolith}.

According to Lewis and Fowler~\cite{lewisfowler14:ms}, \Quote{the
microservice architectural style is an approach to developing a single
application as a suite of small services, each running in its process and
communicating with lightweight mechanisms, often an HTTP resource API.
These services are built around business capabilities and independently
deployable by fully automated deployment machinery. There is a bare minimum of
centralized management of these services, which may be written in different
programming languages and use different data storage technologies.}

In the literature, microservices are often described as an alternative to
monolithic systems that face organizational and technical limitations.
The transition from monolith to microservices can follow the \Pattern{Strangler
Fig} pattern~\cite{fowler04:stranglerfig}, in which existing functionalities
are extracted into independent services until the monolith becomes obsolete and
is replaced entirely. This process is gradual, as a full rewrite of the monolith
is usually infeasible due to its size and complexity.

\subsection{Micro Frontend Architecture}
\label{subsec:micro_frontents_architecture}
%%%%%%%%%%%%%%%%%%%%%%%%%%%%%%%%%%%%%%%%%%%%%%%%%%%%%%%%%%%%%%%%%%%%%%%%%%%%%%%%

Micro frontends are \Quote{an architectural style where independently
deliverable frontend applications are composed into a greater whole}%
~\cite{cam19:mf}. The application is broken down into smaller and simpler
parts that can be developed, tested, and deployed independently,
while preserving a unified user experience.

According to Peltonen \etal~\cite{peltonen21:mvlr},
\Quote{micro-frontends extends the microservice architecture idea and many
principles from microservices apply to micro frontends}.
Therefore, it is common to implement both architectures in parallel.
Since backend and frontend are distinct domains, existing distributed system
knowledge and infrastructure can be leveraged without conflict between services.
For these reasons, micro frontends are considered a natural next step after the
adoption of microservices~\cite{yang19:ramf}.

Peltonen \etal~\cite{peltonen21:mvlr} and Geers~\cite{geers20:mfia} identify
several motivations for adopting a micro frontend architecture. These include
the increasing complexity of frontend systems, the growth of the codebase, and
organizational challenges. Scalability is also a recurring concern, particularly
regarding the need for independent deployments and teams, faster feature
delivery, innovation enablement, and reduced onboarding time for new developers.
In some cases, adoption is also influenced by market trends, with companies
embracing microservices and micro frontends to align with industry best
practices.

Furthermore, micro frontends offer benefits similar to those of microservices.
These include technological flexibility, allowing each part of the system to use
the most appropriate tools, and the ability to form cross-functional teams
focused on specific business domains. The architecture promotes team and project
autonomy, facilitating development, testing, and continuous delivery. It also
supports scalability both at the application and organizational levels. Another
relevant aspect is resilience: failures in a micro frontend affect only that
service, avoiding the full-system outages common in monolithic architectures.

Regarding the challenges of adopting micro frontends, the most significant
include the complexity introduced by the distributed nature of the architecture,
which involves multiple projects and diverse technologies. Ensuring visual
consistency across services is also critical, as the final product must appear
cohesive and uniform to the user. Governance poses another challenge, especially
when different teams share responsibility for the same service or operate within
the same business domain. Finally, performance concerns must be addressed, due
to potential code duplication, shared dependencies, and the volume of data
transferred during interface composition~\cite{peltonen21:mvlr,geers20:mfia}.

Before implementing micro frontends, it is necessary to define which approach
will be adopted. There are two main approaches: the horizontal approach, in
which each micro frontend represents a portion of the page, and the vertical
approach, in which each micro frontend corresponds to a business context or
domain. In the latter, a single micro frontend may contain parts of a page, an
entire page, or even multiple pages~\cite{mezzalira:bmf}.

Another important aspect to consider when implementing a micro frontend
architecture is the adoption of architectural patterns. In this context,
the most relevant patterns are \Pattern{API Gateway} and \Pattern{Backend
for Frontend} (BFF).

The \Pattern{API Gateway} is a service that acts as the entry point for other
services. All requests made by clients -- such as a mobile app or a desktop
website -- go through the gateway, which, in addition to routing them to the
appropriate services, can also handle authorization checks and protocol
translation. This pattern is particularly useful in microservices-based
architectures, where multiple clients access various services. Communication
with each service can occur through different protocols, such as HTTP and gRPC
\cite{richardson18:gateway}.

The \Pattern{Backend for Frontend} (BFF) pattern is a variation of the
\Pattern{API Gateway} pattern~\cite{richardson18:gateway}, in which each client
type has its own dedicated gateway. This allows for the isolation of specific
client needs and enables a team to take responsibility for both the client
application and its corresponding BFF.

%% file: sections/03-architecture_overview.tex
\section{Architecture Overview of \Company}
\label{sec:architecture_overview}
%%%%%%%%%%%%%%%%%%%%%%%%%%%%%%%%%%%%%%%%%%%%%%%%%%%%%%%%%%%%%%%%%%%%%%%%%%%%%%%%

\Company (a fictitious name used for confidentiality) is an online handicraft marketplace founded in 2008. By 2009, it had reached 100,000 products sold and 10,000 active sellers. As of 2022, the platform hosts over 7 million products from more than 100,000 active sellers. Operating in the e-commerce sector, \Company connects independent artisans with customers, enabling the sale of handmade and unique products through its digital marketplace.

The frontend plays a critical role in \Company's success, as it must deliver a performant, user-friendly, and accessible experience across desktop and mobile devices, directly impacting customer satisfaction and retention. Prior to adopting micro frontends, \Company employed the \Pattern{Strangler Fig} pattern to incrementally migrate its legacy monolithic system, referred to here as \Marketplace, to a microservices-based backend architecture. Despite backend progress, the frontend remained centralized within the monolith. Additionally, a separate project, \Aquarelle, was developed to support specific features such as a chat interface between buyers and sellers.

Launched in 2012, \Marketplace handled both backend and frontend responsibilities, serving as the single entry point for all platform functionalities and representing a potential single point of failure. Developed with Java, JSP, Sass, and internal JavaScript libraries, the monolithic design enabled rapid early development but eventually became a bottleneck due to growing complexity, slow deployment cycles, and architectural erosion. Core functionalities like product search, shipping calculations, and category management were progressively extracted into independent microservices following the \Pattern{Strangler Fig} pattern.

The \Aquarelle project, built on Node.js, was introduced to implement a reactive chat feature, capable of displaying dynamic backend data such as order status and user actions. Since modern frontend frameworks were not yet widely adopted in early 2016, the team developed internal libraries to support this functionality. However, \Aquarelle faced several technical and architectural challenges: key internal libraries were deprecated shortly after launch, maintenance suffered from poor documentation, and developer interest waned due to outdated technologies. Furthermore, limited expertise in Node.js and insufficient time to decouple features from the monolith resulted in tight coupling between \Aquarelle and \Marketplace.

\Marketplace also remained responsible for critical cross-cutting concerns such as authentication, data orchestration, and formatting, requiring frontend developers to interact directly with backend code. This rigid architecture, combined with deprecated technologies and organizational constraints, ultimately motivated the adoption of a micro frontend architecture aimed at improving modularity, team autonomy, and deployment agility.

%% file: sections/04-implementing_micro_frontends.tex
\section{Implementing Micro Frontends at \Company}
\label{sec:implementing_micro_frontents}
%%%%%%%%%%%%%%%%%%%%%%%%%%%%%%%%%%%%%%%%%%%%%%%%%%%%%%%%%%%%%%%%%%%%%%%%%%%%%%%%

Due to the issues discussed in the previous section, a plan was devised for the
implementation of a new frontend architecture based on the adoption of modern
technologies. The chosen architecture was micro frontends, which facilitates the
isolation of responsibilities, allows interfaces to be divided into smaller
parts, and enables the use of different technologies. To support this, it was
necessary to create a service that would act as an \Pattern{API Gateway}, so
that the monolith would no longer serve as the central entry point and would be
accessed only when needed. Based on this plan, the projects \Fortelle, \Veloura,
\Nuvelle, and \Brindle (fictitious names) were created, each playing a strategic
role in the construction of the new architecture.

\Fortelle is the implementation of the \Pattern{API Gateway} pattern,
serving as the entry point for \Company's systems. All user requests
pass through it, where routing, validations, and security functions for other
services are handled. This implementation was essential to enable the micro
frontends to start taking over the responsibility of serving user interfaces,
previously centralized in the monolith. In parallel, the team began evaluating
frameworks and libraries to ensure good performance and a positive developer
experience.

\Veloura is \Company's design system, consisting of a set of standards for
building user interfaces -- including colors, spacing, fonts, and icons --
which reinforce the company's visual identity~\cite{hacq:designsystems}.
The \Nuvelle project was conceived as a proof of concept for the selected
frontend technologies. Its development allowed the team to validate practices
such as file organization, supporting libraries, and code conventions.
From \Nuvelle, the first \Brindle was created, with the goal of migrating the
search page for a specific set of craft materials from the monolith to the new
architecture. This page was selected because it was visually similar to the main
search page, but had a lower volume of traffic.

\begin{figure}[ht]
  \centering
  \centering
  \includegraphics[width=0.65\linewidth]{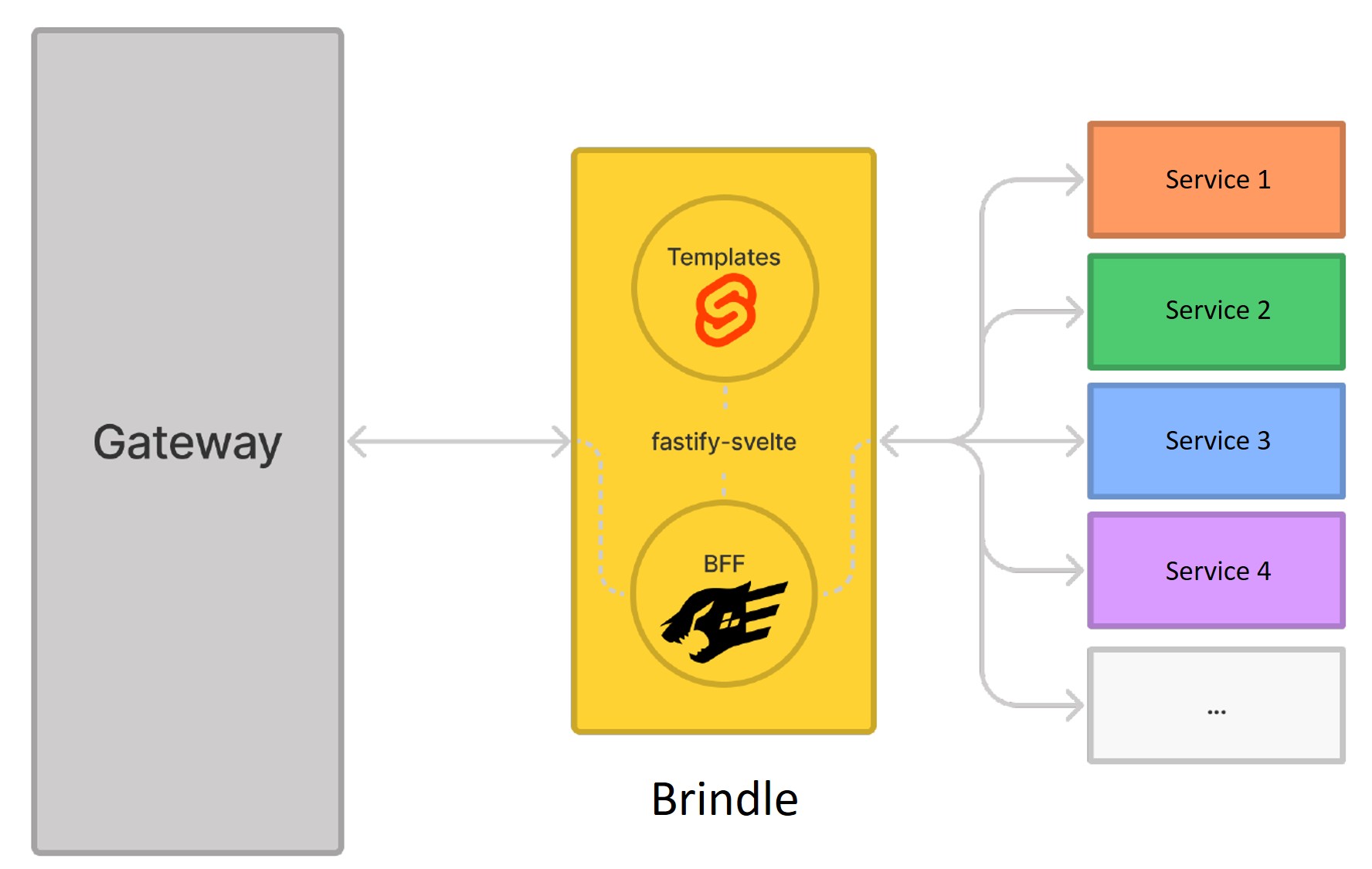}
  \caption{Architectural schematic of a \Brindle project.}
  \label{fig:brindle_architecture}
\end{figure}

After deploying the first \Brindle to production, new clones were developed to
meet internal demands. All systems derived from \Nuvelle came to be known as
\Brindle projects, with additional names used to differentiate them, such as
\BrindleDiscovery, for example. Currently, all of \Company's micro frontends
are \Brindle projects. \autoref{fig:brindle_architecture} illustrates the basic
structure of a micro frontend: user requests pass through the \Pattern{API
Gateway} and are redirected to a \Brindle instance, where the \Pattern{Backend
for Frontend} (BFF) pattern handles internal routing, orchestrates data from
microservices, and forwards it to a template rendered by an open-source library
developed by \Company.

%------------------------------------------------------------------------------%
%------------------------------------------------------------------------------%

With the creation of three \Brindle projects, the frontend teams were able to
identify common modules across services. These shared components were extracted
into a new library, \BrindleCore, an open-source framework for building
\Brindle{s}. One of the main advantages of \BrindleCore is its ability to
abstract internal mechanisms that rarely change, reducing code duplication, the
number of files, and the need for repeated testing in common components.

\break % HACK

%% file: sections/05-current_architecture.tex
\section{Current Architecture of \Company}
\label{sec:current_architecture}
%%%%%%%%%%%%%%%%%%%%%%%%%%%%%%%%%%%%%%%%%%%%%%%%%%%%%%%%%%%%%%%%%%%%%%%%%%%%%%%%

\begin{figure}[ht]
  \centering
  \centering
  \includegraphics[width=0.8\linewidth]{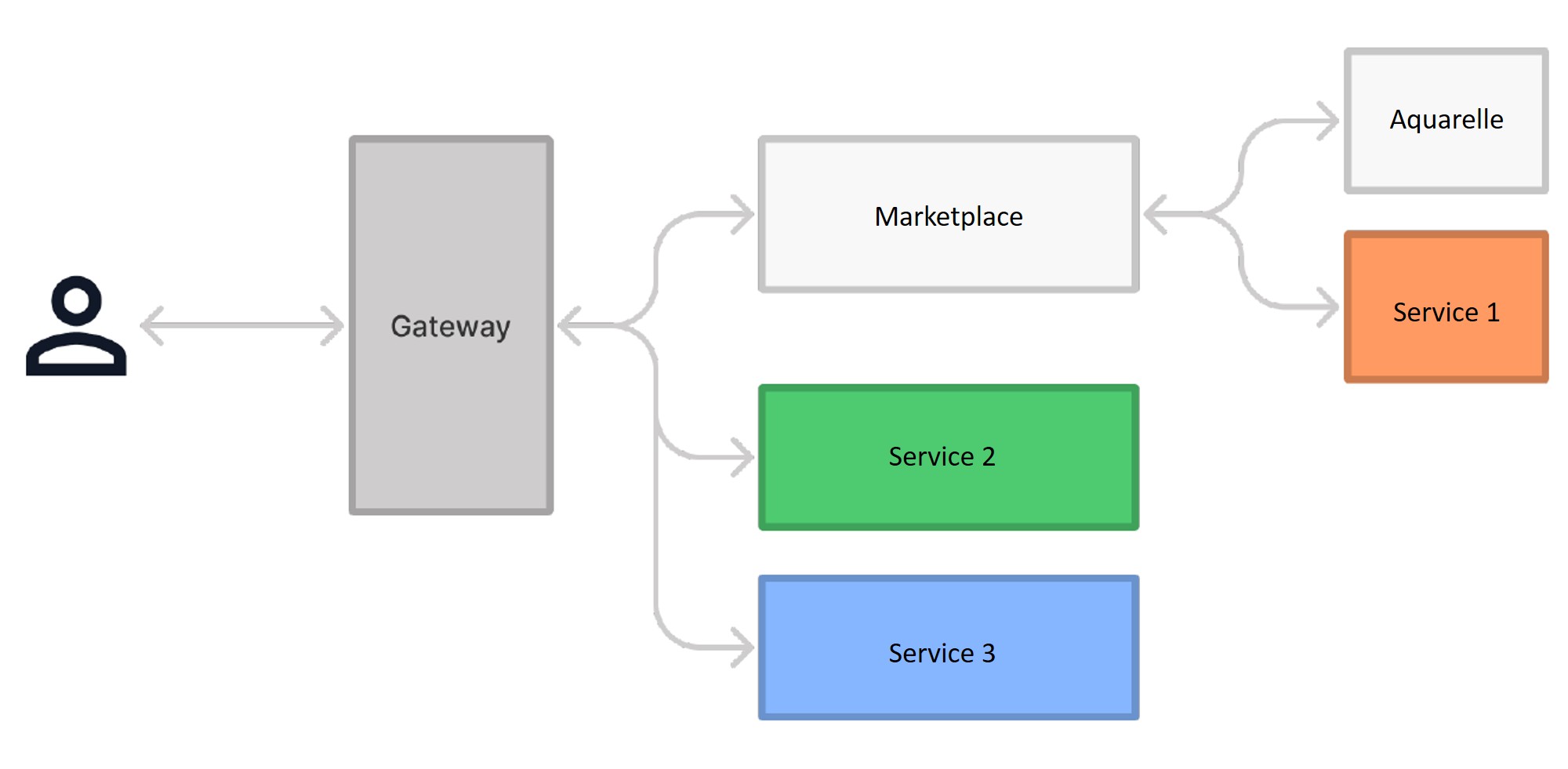}
  \caption{Request flow to frontend after implementing micro frontends.}
  \label{fig:current_architecture}
\end{figure}

With the conception of the mentioned projects, the company's frontend
architecture underwent significant changes.
The introduction of the \Pattern{API Gateway} pattern -- as illustrated by
\autoref{fig:current_architecture} -- enabled other services, besides the
monolith, to handle part of the user request processing, decentralizing
responsibilities and enabling the adoption of micro frontends.
As a result, creating new \Brindle projects became straightforward,
requiring only minor additions to the \Pattern{API Gateway} code.

%------------------------------------------------------------------------------%
%------------------------------------------------------------------------------%

The architectural transition at \Company proved successful. The adoption of
vertical micro frontends allowed the migration of pages from the monolith to the
\Brindle projects, which utilize modern technologies and provide an improved
development experience. Moreover, the new structure facilitated the creation of
frontend systems for the company's new products. It is worth noting that the
ongoing transformation towards microservices favored this adoption by enabling
the reuse of technologies and infrastructure already in place, since micro
frontends also share microservices' principles.

During this research, it was observed that the natural evolution of the
architecture involves migrating more pages from the monolith to the new
services, as well as creating additional micro frontends. However, there
are no plans to expand the number of services, as \Company has defined a
domain division based on its products. Therefore, until new products are
conceived, the teams remain focused on improving existing services.

Nonetheless, a bottleneck exists: to migrate more pages to the \Brindle
projects, functionalities from the monolith must be converted into
microservices. Since the backend teams currently lack the availability to
develop these services, obtaining the necessary data for the interfaces is
compromised. To mitigate this issue, backend, and frontend teams must establish
contracts, allowing simulation of calls to future services and advancing
development. Despite adopting microservices and micro frontends, \Company is
still evaluating the implementation of multidomain teams, considering
organizational impacts and questioning whether this practice is always
beneficial, as discussed in the literature.

%% file: sections/06-survey_with_devs.tex
\section{Survey with Developers}
\label{sec:survey_with_devs}
%%%%%%%%%%%%%%%%%%%%%%%%%%%%%%%%%%%%%%%%%%%%%%%%%%%%%%%%%%%%%%%%%%%%%%%%%%%%%%%%

To better understand the impacts of the architectural change, a semi-open questionnaire was conducted on \Company's employees with knowledge or involvement in frontend projects. This initiative aimed to identify perceived problems, expectations about the change, and observed impacts, particularly regarding the adoption of micro frontends.

The target audience consisted of all active employees during the research period who were directly or indirectly involved in frontend development or decisions. Based on internal information from the engineering team, this included frontend software developers, technical leads, engineering managers, and architecture team members. Invitations were formally sent via institutional email, and the questionnaire was made available through Google Forms for approximately one month, with at least one reminder issued. All eight invited employees responded, yielding a 100\% participation rate. Informed consent was obtained, and responses were anonymized for academic analysis. Among the respondents, only one was a woman. Most participants (62.5\%) had over ten years of experience in the technology field, and half worked specifically as frontend developers.

The questionnaire~\footnote{\scriptsize{The questionnaire is available at: \href{https://drive.google.com/file/d/1yXMi86kuEf7D2dj9hAAEGQx-162-HmIo/view?usp=sharing}{https://drive.google.com/file/d/1yXMi86kuEf7D2dj9hAAEGQx-162-HmIo/view?usp=sharing}}}consisted of 19 questions, 15 open-ended and 4 multiple-choice, designed to distinguish, as much as possible, the impacts of adopting new technologies from those stemming from architectural changes. A Likert scale was used for multiple-choice questions. The questions were organized into five sections: (1) respondent profile (professional experience and involvement in frontend activities), (2) architecture prior to micro frontends, (3) adoption of new technologies, (4) architectural change, and (5) understanding of the new architecture.

Participants were asked to evaluate both technological and architectural impacts across seven key aspects: hiring and onboarding, deployment, development across teams, project understanding, implementation of new features, testability, and development speed. These aspects were selected due to their relevance to the company and frequent recurrence in related literature.

Regarding involvement in frontend activities, whether in development or architectural decisions, respondents rated themselves between medium and high levels. This result was expected, considering that developers are directly involved in implementation while managers typically contribute at a more strategic level.

\begin{figure}[ht]
  \centering
  \centering
  \includegraphics[width=0.8\linewidth]{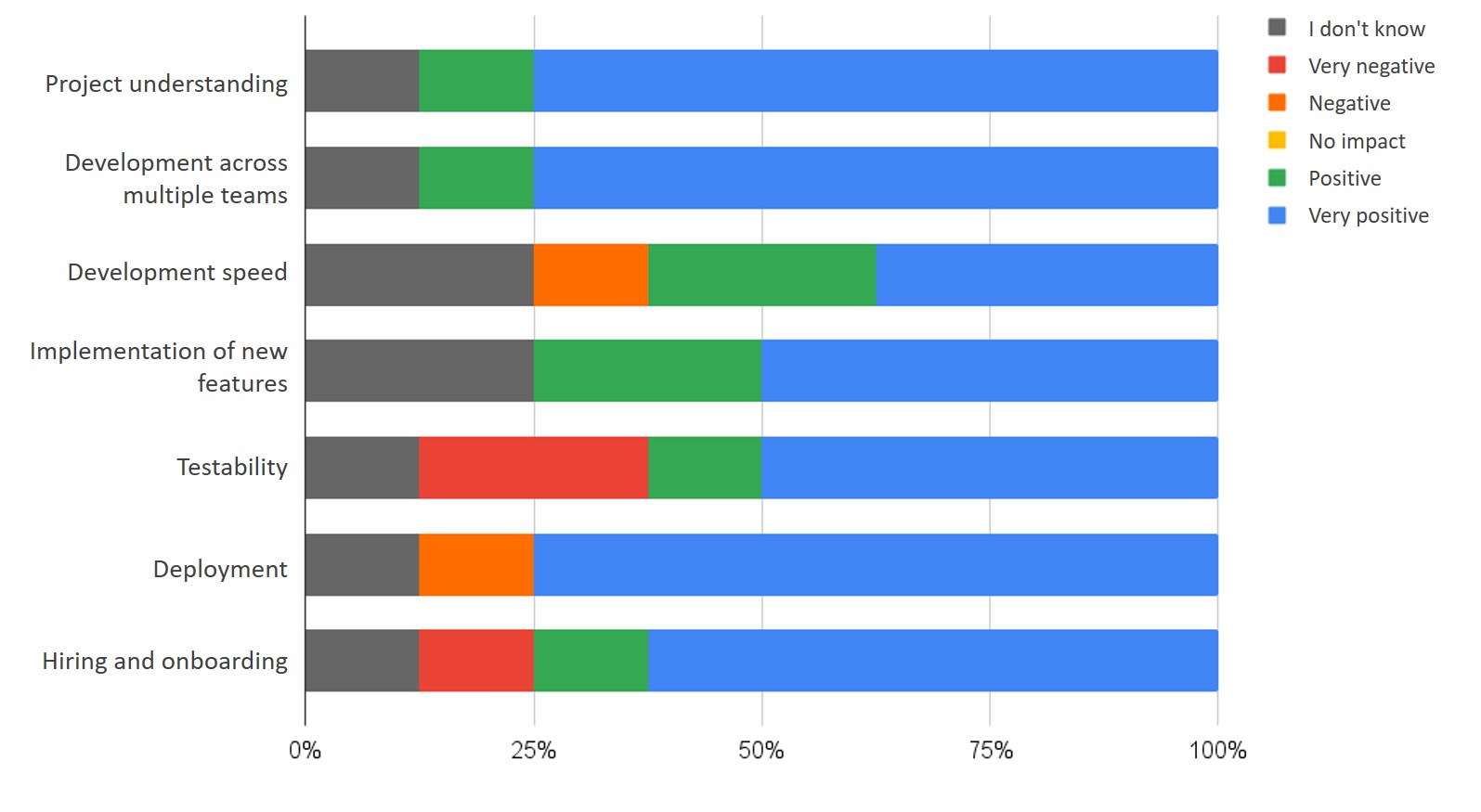}
  \caption{Impacts brought about by new architecture.}
  \label{fig:impacts_new_architecture}
\end{figure}

When asked about the impacts of the new architecture,~\autoref{fig:impacts_new_architecture} shows the emergence of responses such as \Quote{I don’t know} and the lack of responses like \Quote{No impact}, indicating a lack of clarity regarding the architectural effects. Among the predefined aspects, testability and hiring and onboarding processes were reported as the most negatively affected. In terms of testability, although tests were no longer coupled to the monolith, end-to-end testing in distributed systems, such as those based on microservices and micro frontends, proved to be inherently more complex. Regarding hiring and onboarding, the architectural fragmentation enabled a gradual learning path but made it more difficult to understand the system as a whole.

%------------------------------------------------------------------------------%
%------------------------------------------------------------------------------%

Open-ended responses were individually analyzed, and relevant excerpts were coded. These codes were then grouped into three overarching themes: old architecture, new architecture, and micro frontends. Each theme was subdivided to better organize the emerging subtopics.

%\break % HACK

\begin{figure}[ht]
  \centering
  \centering
  \includegraphics[width=0.85\linewidth]{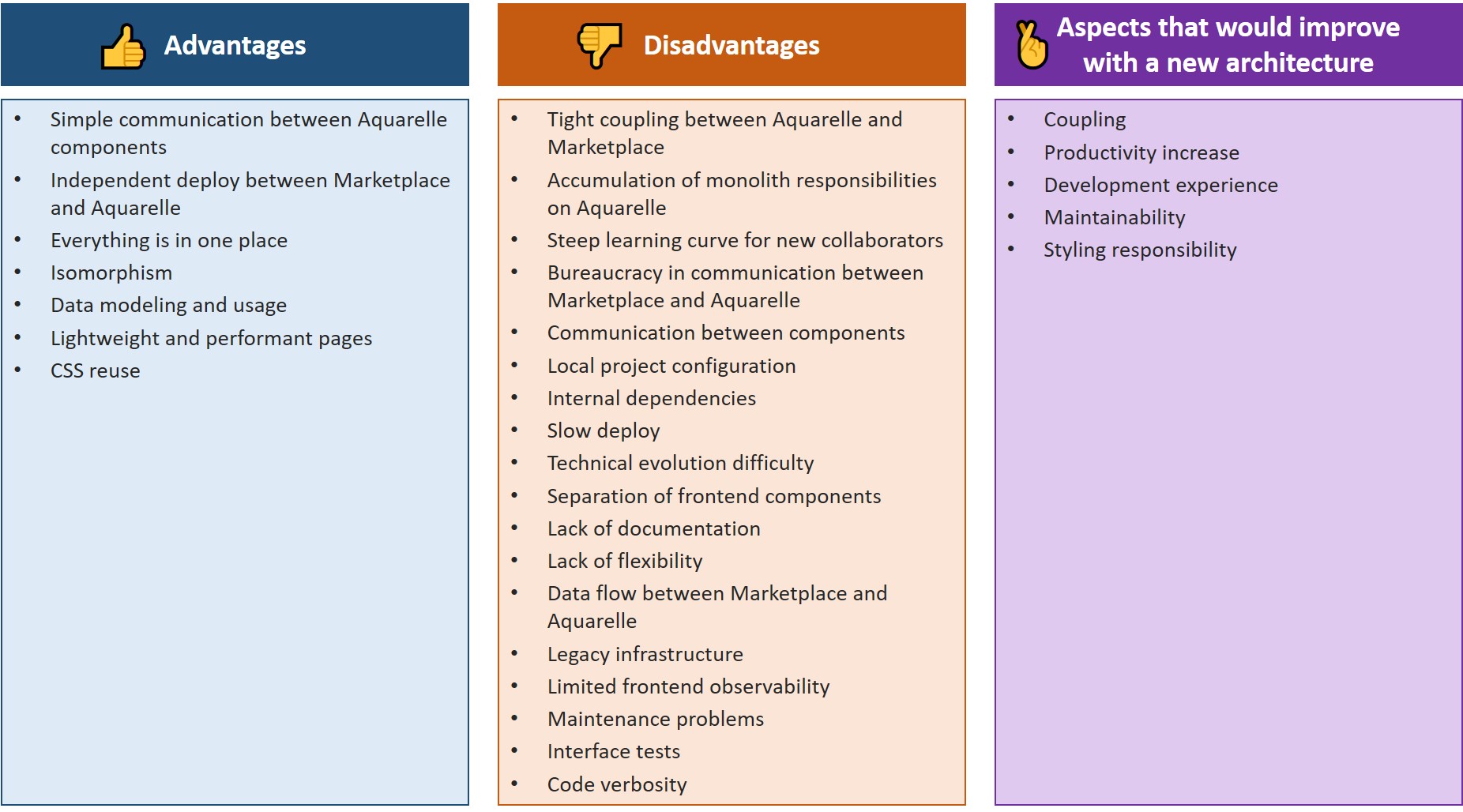}
  \caption{Advantages and disadvantages of old architecture, according to respondents.}
  \label{fig:ad_dis_old_architecture}
\end{figure}

\begin{figure}[ht]
  \centering
  \centering
  \includegraphics[width=0.8\linewidth]{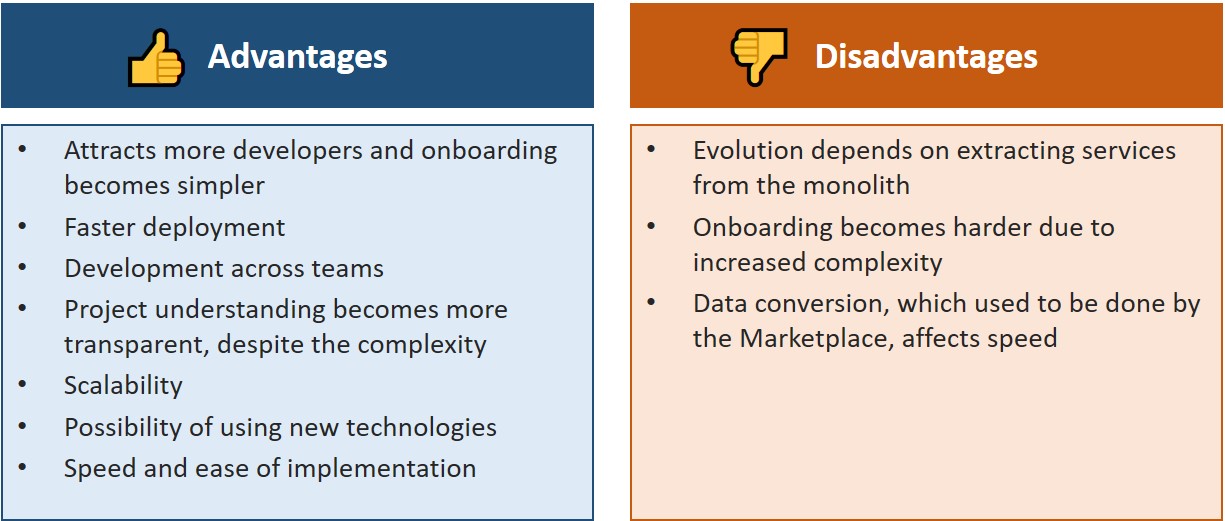}
  \caption{Advantages and disadvantages of new architecture, according to respondents.}
  \label{fig:ad_dis_new_architecture}
\end{figure}

Regarding the old architecture (\autoref{fig:ad_dis_old_architecture}), the analysis focused on identifying its strengths and weaknesses, as well as the expectations for improvement. For the new architecture (\autoref{fig:ad_dis_new_architecture}), the goal was to understand perceived advantages and disadvantages, especially those affecting participants' daily work. Finally, participants were asked to describe their understanding of micro frontends and to evaluate whether the new architecture could be considered a legitimate implementation of that concept.

%------------------------------------------------------------------------------%
%------------------------------------------------------------------------------%

%------------------------------------------------------------------------------%
%------------------------------------------------------------------------------%

Several disadvantages were mentioned. However, some points raised may be more closely related to the technologies used than to the architectural approach itself, as some confusion between technological and architectural impacts was anticipated. Most advantages reported were associated with \Aquarelle, which, despite meeting its objectives, still suffers from technological and architectural limitations. The most frequently cited disadvantage was the strong coupling between frontend systems, precisely the key issue the architectural change aimed to resolve.

Overall, there were more mentions of advantages than disadvantages regarding the new architecture. Some aspects were viewed ambivalently: the use of an emerging architecture helps attract developers, and decomposing the system into smaller services can ease onboarding. Nonetheless, the complexity of distributed systems tends to prolong the onboarding process. Furthermore, although implementation speed increased, it was accompanied by codebase expansion.

\begin{figure}[ht]
  \centering
  \centering
  \includegraphics[width=0.55\linewidth]{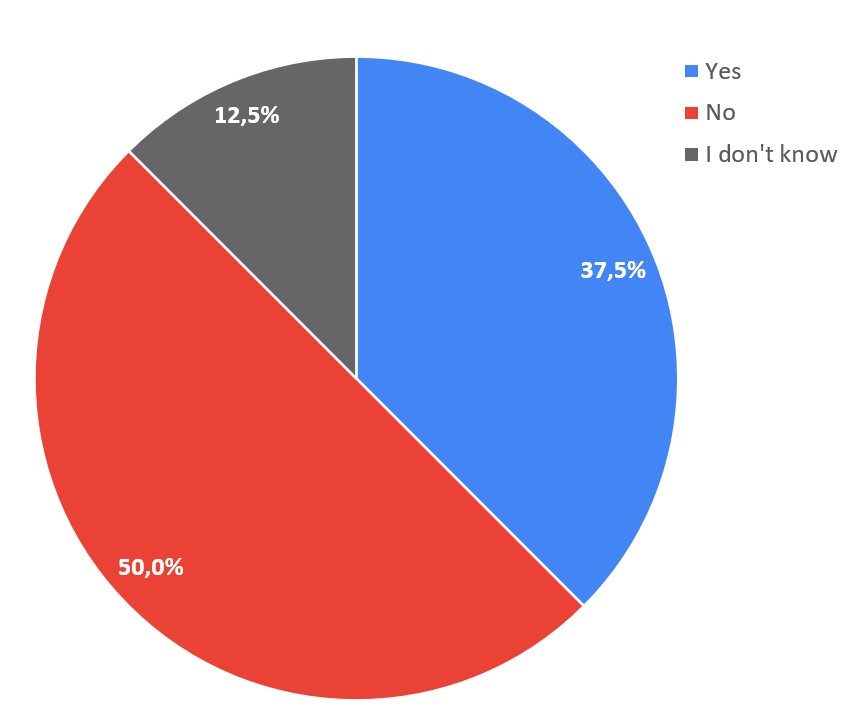}
  \caption{Respondents' perception of whether the new architecture qualifies as a micro frontend.}
  \label{fig:clas_architecture}
\end{figure}

A noteworthy finding is that half of the participants did not recognize the new architecture as a micro frontend implementation (\autoref{fig:clas_architecture}). This perception may stem from two main factors: the dominant definitions in the literature and the absence of explicit references to the vertical slicing approach. Many definitions suggest that a page must be split into multiple visual components to qualify as a micro frontend, which is not necessarily true. As with microservices, there is no strict definition of what constitutes a \emph{service}. Thus, micro frontends may exist even without direct visual interface decomposition.

%------------------------------------------------------------------------------%
%------------------------------------------------------------------------------%

Finally, the analysis indicated that, despite the successful implementation of micro frontends, this was not necessarily the only viable architectural solution for \Company's context. A monolithic frontend, for instance, could have been feasible, provided that decoupling and modern technologies were adequately applied. One finding that supports this argument is that only one participant mentioned scalability, which is commonly cited as one of the main benefits of microservice-based architectures, and by extension, of micro frontends.

%% file: sections/07-related_work.tex
\vspace{-0.1cm}
\section{Related Work}
\label{sec:related-work}
% %%%%%%%%%%%%%%%%%%%%%%%%%%%%%%%%%%%%%%%%%%%%%%%%%%%%%%%%%%%%%%%%%%%%%%%%%%%%%%%%

Some related work to this study has been identified in the literature, investigating relevant aspects of micro frontend architecture adoption and complementing the findings of the present research.

The work by Taibi and Mezzalira~\cite{taibi2022microfrontends} presented the fundamental principles, implementation approaches, and common pitfalls associated with the use of micro frontends. The methodology combined a literature review with practical experience in a sports streaming company. The authors analyzed composition models such as iFrames, Web Components, and Module Federation, as well as communication strategies between micro-applications. The results indicate that, although the architecture fosters team autonomy and scalability in large-scale projects, it also introduces challenges such as increased complexity, the need for robust governance, and significant investments in automation and monitoring. These findings corroborate the technical and organizational challenges identified in this study, especially concerning architectural complexity and governance.

Amorim and Canedo~\cite{amorim2025microfrontend} conducted a systematic mapping study investigating the impacts, challenges, and architectural patterns across 24 scientific studies on micro frontends. Among the most recurrent patterns are Single-SPA, Web Components, Module Federation, and Application Shell. Reported benefits include scalability, flexibility, and efficiency in continuous delivery, while drawbacks involve operational complexity, code redundancy, and difficulties in standardizing user experience. Our study reinforces these results, especially concerning developers’ perceptions of system complexity and challenges in maintaining interface consistency, reflected in onboarding difficulties and conceptual ambiguities.

Sutharsica et al.~\cite{sutharsica2025microfrontend} explored the adoption of micro frontend architecture in startups through a mixed-methods approach combining a systematic literature review, quantitative survey, and qualitative interviews. Their findings reveal that despite common startup constraints such as limited resources, perceived advantages include scalability, modularity, and code reusability. Nevertheless, persistent challenges relate to integration, UI/UX inconsistency, and performance overhead. The emphasis on incremental planning and technical upskilling highlights aspects also observed in our industrial context, where migration required careful alignment among architecture, infrastructure, and teams.

Finally, Peltonen et al.~\cite{peltonen2021motivations} performed a multivocal literature review integrating academic sources and gray literature, focusing on motivations, benefits, and challenges of micro frontend adoption. The results show that adoption is primarily driven by the need to address growing complexity in monolithic frontends and the pursuit of team autonomy and scalability. Key benefits include technological independence and faster feature delivery. On the other hand, challenges such as increased system complexity, user experience inconsistency, code duplication, and governance issues were also identified. These conclusions align with the perceptions reported by developers in our study, who highlighted both advantages related to flexibility and modernization and difficulties concerning complexity, onboarding, and conceptual clarity of the architecture.

Overall, when considering the related work, our study contributes by providing an empirical analysis of micro frontend implementation in a specific industrial setting, emphasizing motivations, challenges, and developer perceptions, and reinforcing the importance of a contextualized understanding for successful adoption of the micro frontend architecture.

%% file: sections/08-conclusion.tex
\vspace{-0.1cm}
\section{Conclusion}
\label{sec:conclusion}
% %%%%%%%%%%%%%%%%%%%%%%%%%%%%%%%%%%%%%%%%%%%%%%%%%%%%%%%%%%%%%%%%%%%%%%%%%%%%%%%%

This paper provides an empirical account of the application of the micro
frontend architectural style in an industry context, with the aim of helping
understand under which circumstances its adoption may be worthwhile. Initially,
a state-of-the-art study on the architecture was conducted through both academic
literature and gray literature sources, such as blog posts and conference talks.
Next, micro frontends were implemented at \Company, and finally, the adoption
was evaluated through surveys with its developers.

The study was guided by two research questions: \textit{\textbf{RQ1}: what are the
motivations and challenges involved in adopting micro frontend architecture in
a company that already uses microservices?} and \textit{\textbf{RQ2}: what are the
perceived benefits and drawbacks reported by developers involved in the
migration?}

The academic literature on micro frontends is still evolving, especially regarding
the concept of vertical, or domain-based, micro frontends as described by Mezzalira%
~\cite{mezzalira:bmf}, whose simplicity facilitates implementation.
Furthermore, existing definitions lack flexibility and generally tie the
architecture to the decomposition of a page into components, which caused
confusion within the company.

Although a vertical approach was adopted, developers hesitated to classify it as
a micro frontend architecture due to the absence of composition, a central focus
in works such as by Geers~\cite{geers20:mfia}. Therefore, there is a need to
align the definition of the architecture more closely with that of its
predecessor, microservices, and to document alternative forms such as micro
frontends without user interfaces.

At \Company, all prerequisites were met, and the adoption was considered
successful. The \Pattern{API Gateway} pattern enabled a distribution of frontend
responsibilities, while the \Pattern{Backend for Frontend} (BFF) pattern
supported communication with backend services. Still, adopting micro frontends
was not strictly necessary, as other architectures, such as a monolithic
frontend, could have achieved comparable results. What made the transition more
feasible was the company's ongoing migration to microservices, which allowed
teams to reuse infrastructure and share existing knowledge.

Regarding \textit{\textbf{RQ1}}, the main motivations for adopting micro frontends included
the need to modernize the frontend stack, overcome the limitations of the monolithic
architecture, and align with the microservices-based backend already in place. Key
challenges involved technical complexity, the coexistence with legacy systems, and
conceptual misunderstandings about the architecture, particularly due to restrictive
definitions found in the literature.

In relation to \textit{\textbf{RQ2}}, developers reported both benefits and drawbacks. Among the
advantages were faster implementation, alignment with modern practices, and greater
flexibility. On the other hand, difficulties included increased system complexity, 
onboarding challenges due to service fragmentation, and ambiguities in defining
whether the solution qualifies as a micro frontend architecture.

Thus, while not the only possible solution, micro frontends turned out to be the
most convenient within that specific context. It is believed that this article may
support researchers and practitioners alike by combining theory and practice, 
offering a historical account of implementation in an e-commerce setting, and 
providing a critical perspective on the literature. Additionally, this study 
contributed to architectural discussions at \Company and serves as documentation
for current and future developers.

%% file: main.bbl
\begin{thebibliography}{10}
\providecommand{\url}[1]{\texttt{#1}}
\providecommand{\urlprefix}{URL }
\providecommand{\doi}[1]{https://doi.org/#1}

\bibitem{amorim2025microfrontend}
Cunha~de Amorim, G., Canedo, E.D.: Micro-frontend architecture in software development: A systematic mapping study. In: Proceedings of the 27th International Conference on Enterprise Information Systems (ICEIS 2025) - Volume 2. pp. 105--116. SCITEPRESS (2025). \doi{10.5220/0013195800003929}

\bibitem{bassclementskazman03:sap}
Bass, L., Clements, P., Kazman, R.: Software Architecture in Practice. Addison Wesley, 3a edn. (2013)

\bibitem{brooks95:mmm}
Brooks, F.: The Mythical Man-Month: Essays on Software Engineering. Addison Wesley (1995)

\bibitem{fowler04:stranglerfig}
Fowler, M.: Stranglerfigapplication. \url{https://martinfowler.com/bliki/StranglerFigApplication.html} (2004), Último acesso em 14/08/2021

\bibitem{geers20:mfia}
Geers, M.: Micro Frontends in Action. Manning, 1a edn. (2020)

\bibitem{hacq:designsystems}
Hacq, A.: Everything you need to know about design systems. \url{https://uxdesign.cc/everything-you-need-to-know-about-design-systems-54b109851969} (2018), Último acesso em 23/12/2021

\bibitem{cam19:mf}
Jackson, C.: Micro frontends. \url{https://martinfowler.com/articles/micro-frontends.html} (Jun 2019), Último acesso em 21/06/2021

\bibitem{lewisfowler14:ms}
Lewis, J., Fowler, M.: Microservices. \url{https://www.martinfowler.com/articles/microservices.html} (Mar 2014), Último acesso em 21/06/2021

\bibitem{martin:ca}
Martin, R.C.: Arquitetura Limpa: O Guia do Artesão para Estrutura e Design de Software. Alta Books Editora (2019)

\bibitem{mezzalira:bmf}
Mezzalira, L.: Building Micro-Frontends. O'Reilley Media, Inc. (2021)

\bibitem{peltonen21:mvlr}
Peltonen, S., Mezzalira, L., Taibi, D.: Motivations, benefits, and issues for adopting micro-frontends: A multivocal literature review. Information and Software Technology  \textbf{136},  106571 (2021). \doi{https://doi.org/10.1016/j.infsof.2021.106571}, \url{https://www.sciencedirect.com/science/article/pii/S0950584921000549}

\bibitem{peltonen2021motivations}
Peltonen, S., Mezzalira, L., Taibi, D.: Motivations, benefits, and issues for adopting micro-frontends: A multivocal literature review. Information and Software Technology  \textbf{136},  106571 (2021). \doi{10.1016/j.infsof.2021.106571}, \url{https://doi.org/10.1016/j.infsof.2021.106571}

\bibitem{richards:sap}
Richards, M.: Software Architecture Patterns. O'Reilley Media, Inc., 1a edn. (2015)

\bibitem{richardson:msp}
Richardson, C.: Microservices Patterns. Manning, 1a edn. (2018)

\bibitem{richardson18:gateway}
Richardson, C.: Pattern: Api gateway / backends for frontends. \url{https://microservices.io/patterns/apigateway.html} (2018), Último acesso em 27/10/2021

\bibitem{richardson19:monolith}
Richardson, C.: Pattern: Monolithic architecture. \url{https://microservices.io/patterns/monolithic.html} (2019), Último acesso em 23/12/2021

\bibitem{sutharsica2025microfrontend}
Sutharsica, A., Arambepola, N.: Micro-frontend architecture: A comparative study of startups and large established companies--suitability, benefits, challenges, and practical insights. In: Proceedings of the 2025 International Conference on Smart Computing and Software Engineering (SCSE). IEEE (2025). \doi{10.1109/SCSE65633.2025.11030972}, \url{https://www.researchgate.net/publication/392684781}

\bibitem{taibi2022microfrontends}
Taibi, D., Mezzalira, L.: Micro-frontends: Principles, implementations, and pitfalls. ACM SIGSOFT Software Engineering Notes  \textbf{47}(4),  25--29 (2022). \doi{10.1145/3561846.3561853}, \url{https://doi.org/10.1145/3561846.3561853}

\bibitem{yang19:ramf}
Yang, C., Liu, C., Su, Z.: Research and application of micro frontends. {IOP} Conference Series: Materials Science and Engineering  \textbf{490},  062082 (Apr 2019). \doi{10.1088/1757-899x/490/6/062082}, \url{https://doi.org/10.1088/1757-899x/490/6/062082}

\end{thebibliography}
